\documentclass[twocolumn,showpacs,preprintnumbers]{revtex4}
\usepackage{bm}
\usepackage{amssymb}
\usepackage{amsmath}
\usepackage{graphicx}
\begin{document}
\title{Constraining URCA cooling of neutron stars \\
       from the neutron radius of $^{208}$Pb}
\author{C.J. Horowitz}\email{horowitz@iucf.indiana.edu}
\affiliation{Department of Physics and Nuclear Theory Center,
             Indiana University, Bloomington, IN 47405}
\author{J. Piekarewicz}\email{jorgep@csit.fsu.edu}
\affiliation{Department of Physics, Florida State University,
             Tallahassee, FL 32306}
\date{\today}
\begin{abstract}
Recent observations by the Chandra observatory suggest that some
neutron stars may cool rapidly, perhaps by the direct URCA process
which requires a high proton fraction. The proton fraction is
determined by the nuclear symmetry energy whose density dependence may
be constrained by measuring the neutron radius of a heavy nucleus,
such as $^{208}$Pb. Such a measurement is necessary for a reliable
extrapolation of the proton fraction to the higher densities present
in the neutron star. A large neutron radius in $^{208}$Pb implies a
stiff symmetry energy that grows rapidly with density, thereby
favoring a high proton fraction and allowing direct URCA
cooling. Predictions for the neutron radius in $^{208}$Pb are
correlated to the proton fraction in dense matter by using a variety
of relativistic effective field-theory models. Models that predict a
neutron ($R_n$) minus proton ($R_p$) root-mean-square radius in
$^{208}$Pb to be $R_n\!-\!R_p\alt0.20$~fm have proton fractions too
small to allow the direct URCA cooling of 1.4~$M_\odot$ neutron
stars. Conversely, if $R_n\!-\!R_p\agt0.25$~fm, the direct URCA
process is allowed (by all models) to cool down a 1.4~$M_\odot$
neutron star. The Parity Radius Experiment at Jefferson Laboratory
aims to measure the neutron radius in $^{208}$Pb accurately and model
independently via parity-violating electron scattering. Such a
measurement would greatly enhance our ability to either confirm or
dismiss the direct URCA cooling of neutron stars.
\end{abstract}
\smallskip
\maketitle

\section{Introduction}

Neutron stars are created with very high temperatures in supernova 
explosions. Indeed, neutrinos observed from SN1987A indicate a 
neutrinosphere temperature that could be as high as 5 MeV~\cite{raf}.  
Neutrons stars then cool, primarily by neutrino emission~\cite{Pe92}. 
In the standard scenario, the modified URCA reaction,
\begin{equation}
n + n \rightarrow n + p + e^- + \bar\nu_{e},
\label{modified}
\end{equation}
emits neutrinos from the volume of the star. This process, however,
is relatively slow as a second nucleon is necessary to conserve 
momentum.

Recent X-ray observations of the neutron star in 3C58~\cite{3c58},
Vela~\cite{vela}, and Geminga~\cite{geminga} indicate low surface
temperatures. Moreover, the low quiescent luminosity in the
transiently accreting binaries KS 1731-260~\cite{wijnands} and Cen
X-4~\cite{page} suggest rapid cooling. As X-ray observatories progress
and our knowledge of neutron-star atmospheres and ages improve,
additional ``cold'' neutron stars may be discovered. Such low surface
temperatures appear to require enhanced cooling from reactions that
proceed faster than the modified URCA process of
Eq.~(\ref{modified}). Enhanced cooling may occur via the weak decay 
of additional hadrons such as pion or kaon condensates~\cite{kaon},
hyperons~\cite{Pr92}, or quark matter~\cite{quarks}. Yet perhaps the most
conservative enhanced-cooling mechanism is the direct URCA
process~\cite{urca,leinson} of neutron beta decay followed by 
electron capture:
\begin{subequations}
 \begin{eqnarray}
  && n \rightarrow p + e^{-} + \bar\nu_{e}\;, \label{urca1}\\
  && e^{-} + p \rightarrow n + \nu_{e}\;.     \label{urca2}
 \end{eqnarray}
\end{subequations}
This mechanism is not ``exotic'' as it only requires protons,
neutrons, and electrons---constituents known to be present in
dense matter. However, to conserve momentum in Eq.~(\ref{urca1}) 
the sum of the Fermi momenta of the protons plus that of
the electrons must be greater than (or equal to) the neutron 
Fermi momentum. This requires a relatively large proton fraction.

Yakovlev and collaborators~\cite{yakovlev} are able to reproduce
measured neutron-star temperatures using a relativistic mean-field
equation of state that allows direct URCA for neutron stars with
masses above $1.358~M_\odot$ ($M_{\odot}\!=\!$~solar mass). In
contrast, Tsuruta and collaborators~\cite{t5} rely on pion
condensation to reproduce the measured temperatures. They argue that
microscopic calculations of neutron-rich matter~\cite{fp} using
nonrelativistic nucleon-nucleon interactions yield too small a proton
fraction for the URCA process to operate. Unfortunately, these
microscopic calculations depend on a poorly known three-nucleon force
and on relativistic effects that could end up increasing the proton
fraction at high densities.

Superconductivity and superfluidity can greatly influence neutron-star
cooling~\cite{Ya99,Pa00}.  For temperatures much lower than the
pairing gap, pairing correlations suppress exponentially the rate of
many cooling reactions. Yet for temperatures of the order of the
pairing gap, the thermal breaking and subsequent reformation of
nucleon ``Cooper'' pairs promotes an additional neutrino-emission
mechanism that rapidly cools the star~\cite{pairing}. However, it has
been argued in Ref.~\cite{yakovlev} that this mechanism alone is
unlikely to explain the low temperature of some neutron stars. This is
because for a large enough neutron-pairing gap, pair breaking would
rapidly cool all neutron stars at a rate almost independent of the
mass of the star.  This would disagree with observations of some warm
neutron stars.  Tsuruta and collaborators have claimed that
microscopic calculations with a high proton concentration show a small
proton pairing gap~\cite{tpairing}. If so, a direct URCA process (one
not controlled by pairing correlations) will cool a star so quickly
that thermal radiation would become invisible~\cite{t5}. However, we
caution that drawing definitive conclusions from microscopic
calculations of pairing gaps may be premature, as significant
uncertainties remain in the interactions, equation of state,
composition, and phases of high-density matter.

Although the precise mechanism remains unknown, some kind of enhanced 
cooling appears to be required to explain the recent observations 
of cold neutron stars. While the need for exotic phases of matter
is appealing, more conventional cooling scenarios, such as the 
direct URCA process, can not be dismissed on purely theoretical
grounds. Moreover, neutron-star observations alone may not be able 
to resolve the detailed mechanism of enhanced cooling. Thus, we 
consider complementary laboratory experiments that could help us
confirm (or possibly dismiss) the direct URCA process. This can be 
achieved by constraining the symmetry energy of dense matter. The 
symmetry energy describes how the energy of (asymmetric) nuclear 
matter increases as one departs from equal numbers of neutrons and 
protons. The proton fraction $Y_{p}\!=\!Z/A$ of nuclear matter in
beta-equilibrium is sensitive to the symmetry energy~\cite{urca}.  
A large symmetry energy imposes a stiff penalty on the system for 
upsetting the $N=Z$ balance hereby forcing it to retain a large 
proton fraction.

Energetic heavy-ion collisions probe the symmetry energy at high
nuclear densities~\cite{li}. Possible observables include the ratio of
$\pi^-$-to-$\pi^+$ production and the neutron-proton differential
collective flow. However, these reactions may suffer from important
uncertainties associated with the complex strong interactions of the
heavy-ion collisions. Thus, we rely on a purely electroweak reaction
that can be unambiguously interpreted. Parity violating elastic
electron scattering from a heavy nucleus is sensitive to the neutron
density. This is because the weak charge of a neutron is much larger
than the weak charge of a proton. The Parity Radius Experiment at the
Jefferson Laboratory aims to measure the neutron radius in $^{208}$Pb
to a 1\% accuracy ($\pm 0.05$~fm)~\cite{prex}. This measurement can be
both accurate and model independent~\cite{bigpaper}. In
Ref.~\cite{brown} Brown showed that the neutron radius of $^{208}$Pb
determines the pressure of neutron-rich matter at normal densities
which, in turn, is related to the density dependence of the symmetry
energy~\cite{prakash}. In an earlier work we showed how the neutron
radius of $^{208}$Pb determines properties of the neutron-star
surface, such as the transition density from a solid crust to a liquid
interior~\cite{prl}. Furthermore, we argued that by comparing the
neutron radius of $^{208}$Pb (a low-density observable) to the radius
of a neutron star (a high and low-density observable) evidence may be
provided in support of a phase transition in dense
matter~\cite{radii}.

In the present work we show how the neutron radius of a heavy nucleus
(such as $^{208}$Pb) controls the density dependence of the symmetry
energy. Unfortunately, the density dependence of the symmetry energy
($da_{\rm sym}/d\rho$) is poorly known.  Thus a measurement of the
neutron radius of $^{208}$Pb seems vital, as it will constrain the
density dependence of the symmetry energy at low density. This, in
turn, will allow a more reliable extrapolation of the symmetry energy,
and thus a more reliable determination of the proton fraction at the
higher densities required in the study of neutron-star structure.
While in principle collective modes of nuclei, such as the
giant-dipole or isovector-monopole resonances, are sensitive to
$da_{\rm sym}/d\rho$, in practice this sensitivity is small.
Moreover, the parameter sets used in the calculations (see various
tables) have been adjusted so that well known ground-state properties
remain fixed while changing the neutron radius. This shows that
existing ground-state information, such as charge densities or binding
energies, do not determine the neutron radius uniquely. Thus the need
for a new measurement---such as the neutron radius in
$^{208}$Pb---that will provide important information on 
$da_{\rm sym}/d\rho$.

The paper has been organized as follows. In Sec.~\ref{sec:formalism},
relativistic effective-field theories for both dense matter and finite
nuclei are discussed. A large number of parameter sets are considered
so that the density-dependence of the symmetry energy may be changed
while reproducing existing ground-state data. In
Sec.~\ref{sec:results}, results for the equilibrium proton fraction as
a function of baryon density are presented using interactions that
predict different neutron radii in $^{208}$Pb. Our summary and
conclusions are offered in Sec.~\ref{sec:conclusions}. In particular,
we conclude that for models with a large neutron skin in $^{208}$Pb
($R_{n}\!-\!R_{p}\agt0.25$~fm) the symmetry energy rises rapidly with
density and the direct URCA cooling of a $1.4~M_\odot$ neutron star is
likely. Conversely, if the neutron radius is small
($R_{n}\!-\!R_{p}\alt 0.20$~fm) it is unlikely that the direct URCA
process occurs. In this case, the enhanced cooling of neutron stars
may indeed require the presence of exotic states of matter, such as
meson condensates, hyperonic, and/or quark matter.

\section{Formalism}
\label{sec:formalism}
Our starting point will be the relativistic effective-field theory 
of Ref.~\cite{horst} supplemented with additional couplings between 
the isoscalar and the isovector mesons. This allows us to correlate
nuclear observables sensitive to the density dependence of the
symmetry energy, such as the neutron radius of $^{208}$Pb, with
neutron-star properties, such as the threshold mass for URCA
cooling. As the density dependence of the symmetry energy is poorly
known, uncertainties in these correlations will be explored by
considering a wide range of model parameters. The interacting
Lagrangian density is thus given by~\cite{prl,horst}
\begin{eqnarray}
{\mathcal L}_{\rm int} &=&
\bar\psi \left[g_{\rm s}\phi   \!-\! 
         \left(g_{\rm v}V_\mu  \!+\!
    \frac{g_{\rho}}{2}{\mbox{\boldmath $\tau$}}\cdot{\bf b}_{\mu} 
                               \!+\!    
    \frac{e}{2}(1\!+\!\tau_{3})A_{\mu}\right)\gamma^{\mu}
         \right]\psi \nonumber \\
                   &-& 
    \frac{\kappa}{3!} (g_{\rm s}\phi)^3 \!-\!
    \frac{\lambda}{4!}(g_{\rm s}\phi)^4 \!+\!
    \frac{\zeta}{4!}   g_{\rm v}^4(V_{\mu}V^\mu)^2 
    \nonumber \\
                   &+&
    g_{\rho}^{2}\,{\bf b}_{\mu}\cdot{\bf b}^{\mu}
    \left[\Lambda_{\rm s} g_{\rm s}^{2}\phi^2 +
          \Lambda_{\rm v} g_{\rm v}^{2}V_{\mu}V^\mu\right] \;.
 \label{LDensity}
\end{eqnarray}
The model contains an isodoublet nucleon field ($\psi$) interacting
via the exchange of two isoscalar mesons, the scalar sigma ($\phi$)
and the vector omega ($V^{\mu}$), one isovector meson, the rho (${\bf
b}^{\mu}$), and the photon ($A^{\mu}$). In addition to meson-nucleon
interactions the Lagrangian density includes scalar and vector
self-interactions. The scalar-meson self-interactions ($\kappa$ and
$\lambda$) soften the equation of state (EOS) of symmetric nuclear
matter at (and near) saturation density while the $\omega$-meson
self-interactions ($\zeta$) soften the high-density component of the
EOS.  Finally, the nonlinear couplings ($\Lambda_{\rm s}$ and
$\Lambda_{\rm v}$) are included to modify the density dependence of
the symmetry energy~\cite{prl,radii}.

The energy of neutron-rich matter may be written in terms of the 
energy of symmetric nuclear matter ($\rho_p\!=\!\rho_n$) and the 
symmetry energy $a_{\rm sym}(\rho)$. That is,
\begin{equation}
  \frac{E}{A}(\rho,t) = \frac{E}{A}(\rho,t=0) 
	              + t^{2} a_{\rm sym}(\rho) 
                      + {\cal O}(t^{4}) \;,
 \label{EOS}
\end{equation}     
where the neutron excess has been defined as
\begin{equation}
    t  \equiv \frac{\rho_{n} - \rho_{p}}
                   {\rho_{n} + \rho_{p}} \;.
 \label{tDef}
\end{equation}     
Here $\rho_n$ is the neutron and $\rho_p$ the proton density, and
\begin{equation}
  \rho=\rho_p+\rho_n\equiv\frac{2k_{\rm F}^{3}}{3\pi^{2}} \;.
 \label{Rhob}
\end{equation}     
The symmetry energy describes how the energy of the system increases 
as one moves away from $\rho_p\!=\!\rho_n$. It is discussed in 
Ref.~\cite{radii} where it is shown that it is given by
\begin{equation}
  a_{\rm sym}(\rho) = \frac{k_{F}^{2}}{6E_{F}^{*}} 
          + \frac{g_{\rho}^{2}}{12\pi^{2}}
            \frac{k_{F}^{3}}{m_{\rho}^{*2}} \;,
 \label{SymmE}
\end{equation}   
where $k_{\rm F}$ is the Fermi momentum, 
$E_{F}^{*}\!=\!\sqrt{k_{\rm F}^{2}+M^{*2}}$, 
and $M^*\!=\!M\!-\!g_s\phi_0$ is the effective 
nucleon mass. Further, the effective rho-meson 
mass has been defined as follows:
\begin{equation}
  m_{\rho}^{*2} = m_{\rho}^{2} + 2g_{\rho}^{2}
  \Big(\Lambda_{\rm s} g_{\rm s}^2\phi_{0}^2 +
       \Lambda_{\rm v} g_{\rm v}^2V_{0}^{2} \Big) \;.
 \label{Mrho2}
\end{equation}     
The symmetry energy is given as a sum of two contributions. The first
term in Eq.~(\ref{SymmE}) represents the increase in the kinetic
energy of the system due to the displacement of the Fermi levels of
the two species (neutrons and protons). This contribution has been
fixed by the properties of symmetric nuclear matter as it only depends
on the nucleon effective mass $M^{*}$. By itself, it leads to an
unrealistically low value for the symmetry energy; for example, at
saturation density this contribution yields $\sim\!15$~MeV, rather
than the most realistic value of $\sim\!37$~MeV. The second
contribution is due to the coupling of the rho meson to an
isovector-vector current that no longer vanishes in the $N\!\ne\!Z$
system. It is by adjusting the strength of the $NN\rho$ coupling
constant that one can now fit the empirical value of the symmetry
energy at saturation density. However, the symmetry energy at
saturation is not well constrained experimentally. Yet an average of
the symmetry energy at saturation density and the surface symmetry
energy is constrained by the binding energy of nuclei.  Thus, the
following prescription is adopted: the value of the $NN\rho$ coupling
constant is adjusted so that all parameter sets have a symmetry energy
of $a_{\rm sym}\!=\!25.67$~MeV at 
$k_F\!=\!1.15$~fm$^{-1}$($\rho\!=\!0.10$~fm$^{-3}$)~\cite{prl}. 
That is,
\begin{equation}
  g_{\rho}^{2}= \frac{m_{\rho}^{2}\,\Delta a_{\rm sym}}
    {\displaystyle{\frac{k_F^3}{12\pi^{2}}}
     -2\left(\Lambda_{\rm s} g_{\rm s}^2\phi_{0}^2 +
       \Lambda_{\rm v} g_{\rm v}^2V_{0}^{2} \right)
       \Delta a_{\rm sym}}\;,
 \label{grho2}
\end{equation}
where 
$\Delta a_{\rm sym}\!\equiv\!(a_{\rm sym}\!-\!k_{F}^{2}/6E_{F}^{*})$. 
This prescription insures accurate binding energies for heavy nuclei,
such as $^{208}$Pb. Following this prescription the symmetry energy at 
saturation density is predicted  
(for $\Lambda_{\rm s}\!=\!\Lambda_{\rm v}\!=\!0$) to be $37.3$, $36.6$, 
and $36.3$~MeV for the three families of parameter sets considered 
in this work, namely NL3~\cite{NL3}, S271~\cite{prl}, 
and Z271\cite{prl}, respectively (see various tables).
Moreover, all these parameter sets reproduce the following properties 
of symmetric nuclear matter: {\it i)} nuclear saturation at a Fermi 
momentum of $k_{\rm F}\!=\!1.30$~fm$^{-1}$ with {\it ii)} a binding 
energy per nucleon of $16.24$ MeV, and {\it iii)} a nuclear 
incompressibility of $K\!=\!271$~MeV. These values follow from the 
successful parametrization of Ref.~\cite{NL3} and, thus, have been 
adopted for the other sets (S271 and Z271) as well. Yet the various 
parameter sets differ in {\it i)} their values for the effective
nucleon mass at saturation density, {\it ii)} the value of the
$\omega$-meson quartic coupling ($\zeta\!\ne\!0$ for set Z271 but
vanishes for the NL3 and S271 sets) and {\it iii)} the nonlinear
couplings $\Lambda_{\rm s}$ and $\Lambda_{\rm v}$ (see various
tables). Note that changing $\Lambda_{\rm s}$ or $\Lambda_{\rm v}$
modifies the density dependence of the symmetry energy through a
change in the effective rho-meson mass [see Eq.~(\ref{Mrho2})]. In
general, increasing either $\Lambda_{\rm s}$ or $\Lambda_{\rm v}$
causes the symmetry energy to soften, that is, to grow slower with
increasing density. This, in turn, allows for a larger neutron-proton
mismatch or equivalently, for a lower equilibrium proton fraction at
high density.
\begin{table}
\caption{Model parameters used in the calculations. The 
parameter $\kappa$ and the scalar mass $m_{\rm s}$ are 
given in MeV. The nucleon, rho, and omega masses are kept 
fixed at $M\!=\!939$, $m_{\rho}\!=\!763$, and 
$m_{\omega}\!=\!783$~MeV, respectively---except in the 
case of the NL3 model where it is fixed at 
$m_{\omega}\!=\!782.5$~MeV.}
\begin{ruledtabular}
\begin{tabular}{lcccccc}
 Model & $m_{\rm s}$  & $g_{\rm s}^2$ & $g_{\rm v}^2$ & 
           $\kappa$ & $\lambda$ & $\zeta$ \\
 \hline
 NL3  & 508.194 & 104.3871  & 165.5854 
      & 3.8599 & $-$0.01591 &   0.00 \\ 
 S271 & 505.000 &  81.1071  & 116.7655 
      & 6.6834 & $-$0.01580 &   0.00 \\ 
 Z271 & 465.000 &  49.4401  &  70.6689 
      & 6.1696 & $+$0.15634 &   \ 0.06 
\label{Table1}
\end{tabular}
\end{ruledtabular}
\end{table}
The neutron radius of $^{208}$Pb also depends on the density
dependence of the symmetry energy. A stiff density dependence ({\it
i.e.,} pressure) for neutron matter pushes neutrons out against
surface tension, leading to a large neutron radius. The pressure of
neutron-rich matter depends on the derivative of the energy of
symmetric matter with respect to the density
($dE(\rho,t\!=\!0)/d\rho$) and on the derivative of the symmetry
energy ($da_{\rm sym}/d\rho$). While the former is well known, at
least in the vicinity of the saturation density, the latter is
not. Hence, by changing the values of $\Lambda_{\rm s}$ or
$\Lambda_{\rm v}$ one can adjust the density dependence of the
symmetry energy $da_{\rm sym}/d\rho$, while keeping a variety of
well-known ground-state properties (such as the proton radius and the
binding energy of $^{208}$Pb) unchanged. Note that parameter sets with
a large ``pressure'', $da_{\rm sym}/d\rho$, yield a large neutron
radius in $^{208}$Pb.

\begin{table}
\caption{Results for the NL3 parameter set with 
	 $\Lambda_{\rm s}~=~0$.
         The neutron skin ($R_{n}\!-\!R_{p}$) of $^{208}$Pb is given
         along with the threshold density for the direct URCA process
         $\rho_{\scriptscriptstyle{\rm URCA}}$, the corresponding
	 proton fraction $Y_{p\,{\rm URCA}}$, and the minimum mass
         neutron star where the direct URCA process is allowed
         $M_{\rm URCA}$. Note that $R_{n}\!-\!R_{p}$ is given in
         fm, $\rho_{\scriptscriptstyle{\rm URCA}}$ in fm$^{-3}$,
         and $M_{\rm URCA}$ in solar masses.}
\begin{ruledtabular}
\begin{tabular}{cccccc}
$\Lambda_{\rm v}$ & $g_\rho^2$ & $R_n\!-\!R_p$ &
$\rho_{\scriptscriptstyle{\rm URCA}}$ & 
$Y_{p\,{\rm URCA}}$ & $M_{\rm URCA}$ \\
\hline
0.0000 &  79.6 & 0.280 & 0.205 & 0.130 &   0.838 \\
0.0050 &  84.9 & 0.266 & 0.233 & 0.131 &   0.944 \\
0.0100 &  90.9 & 0.251 & 0.271 & 0.132 &   1.224 \\
0.0125 &  94.2 & 0.244 & 0.293 & 0.133 &   1.435 \\
0.0150 &  97.9 & 0.237 & 0.319 & 0.134 &   1.671 \\
0.0200 & 106.0 & 0.223 & 0.376 & 0.135 &   2.123 \\
0.0250 & 115.6 & 0.209 & 0.442 & 0.136 & \ 2.449
\label{Table2}
\end{tabular}
\end{ruledtabular}
\end{table}
\begin{table}
\caption{Results for the S271 parameter set with 
	 $\Lambda_{\rm s}~=~0$.  
         The neutron skin ($R_{n}\!-\!R_{p}$) of $^{208}$Pb is given
         along with the threshold density for the direct URCA process
         $\rho_{\scriptscriptstyle{\rm URCA}}$, the corresponding
	 proton fraction $Y_{p\,{\rm URCA}}$, and the minimum mass
         neutron star where the direct URCA process is allowed
         $M_{\rm URCA}$. Note that $R_{n}\!-\!R_{p}$ is given in
         fm, $\rho_{\scriptscriptstyle{\rm URCA}}$ in fm$^{-3}$,
         and $M_{\rm URCA}$ in solar masses.}
\begin{ruledtabular}
\begin{tabular}{cccccc}
$\Lambda_{\rm v}$ & $g_\rho^2$ & $R_n\!-\!R_p$ &
$\rho_{\scriptscriptstyle{\rm URCA}}$ & 
$Y_{p\,{\rm URCA}}$ & $M_{\rm URCA}$ \\
\hline
0.000 &  85.4357 & 0.254 & 0.224 & 0.130 &   0.830 \\
0.005 &  88.3668 & 0.246 & 0.252 & 0.132 &   0.894 \\
0.010 &  91.5061 & 0.238 & 0.296 & 0.133 &   1.059 \\
0.015 &  94.8767 & 0.230 & 0.374 & 0.135 &   1.429 \\
0.020 &  98.5051 & 0.221 & 0.501 & 0.137 &   1.938 \\
0.025 & 102.4221 & 0.214 & 0.663 & 0.139 &   2.248 \\
0.030 & 106.6635 & 0.205 & 0.843 & 0.140 & \ 2.343
\label{Table3}
\end{tabular}
\end{ruledtabular}
\end{table}
\begin{table}
\caption{Results for the Z271 parameter set with 
	 $\Lambda_{\rm s}~=~0$.  
         The neutron skin ($R_{n}\!-\!R_{p}$) of $^{208}$Pb is given
         along with the threshold density for the direct URCA process
         $\rho_{\scriptscriptstyle{\rm URCA}}$, the corresponding
	 proton fraction $Y_{p\,{\rm URCA}}$, and the minimum mass
         neutron star where the direct URCA process is allowed
         $M_{\rm URCA}$. Note that $R_{n}\!-\!R_{p}$ is given in
         fm, $\rho_{\scriptscriptstyle{\rm URCA}}$ in fm$^{-3}$,
         and $M_{\rm URCA}$ in solar masses.}
\begin{ruledtabular}
\begin{tabular}{cccccc}
$\Lambda_{\rm v}$ & $g_\rho^2$ & $R_n\!-\!R_p$ &
$\rho_{\scriptscriptstyle{\rm URCA}}$ & 
$Y_{p\,{\rm URCA}}$ & $M_{\rm URCA}$ \\
\hline
0.000 &  90.2110 & 0.241 & 0.242 & 0.131 &   0.816 \\
0.010 &  92.5415 & 0.235 & 0.274 & 0.132 &   0.862 \\
0.020 &  94.9956 & 0.228 & 0.332 & 0.134 &   0.971 \\
0.025 &  96.2721 & 0.225 & 0.386 & 0.135 &   1.079 \\
0.030 &  97.5834 & 0.222 & 0.500 & 0.137 &   1.270 \\
0.035 &  98.9310 & 0.219 & 0.747 & 0.139 &   1.498 \\
0.040 & 100.3162 & 0.215 & 1.028 & 0.141 & \ 1.583
\label{Table4}
\end{tabular}
\end{ruledtabular}
\end{table}
\begin{table}
\caption{Results for the Z271 parameter set with 
	 $\Lambda_{\rm v}~=~0$.  
         The neutron skin ($R_{n}\!-\!R_{p}$) of $^{208}$Pb is given
         along with the threshold density for the direct URCA process
         $\rho_{\scriptscriptstyle{\rm URCA}}$, the corresponding
	 proton fraction $Y_{p\,{\rm URCA}}$, and the minimum mass
         neutron star where the direct URCA process is allowed
         $M_{\rm URCA}$. Note that $R_{n}\!-\!R_{p}$ is given in
         fm, $\rho_{\scriptscriptstyle{\rm URCA}}$ in fm$^{-3}$,
         and $M_{\rm URCA}$ in solar masses.}
\begin{ruledtabular}
\begin{tabular}{cccccc}
$\Lambda_{\rm s}$ & $g_\rho^2$ & $R_n\!-\!R_p$ &
$\rho_{\scriptscriptstyle{\rm URCA}}$ & 
$Y_{p\,{\rm URCA}}$ & $M_{\rm URCA}$ \\
\hline
0.000 &  90.2110 & 0.241 & 0.242 & 0.131 &   0.816 \\
0.010 &  96.3974 & 0.229 & 0.287 & 0.133 &   0.901 \\
0.020 & 103.4949 & 0.216 & 0.366 & 0.135 &   1.078 \\
0.030 & 111.7205 & 0.204 & 0.488 & 0.137 &   1.300 \\
0.040 & 121.3666 & 0.191 & 0.636 & 0.139 &   1.467 \\
0.050 & 132.8358 & 0.178 & 0.789 & 0.140 &   1.560 \\
0.060 & 146.6988 & 0.164 & 0.936 & 0.141 & \ 1.605
\label{Table5}
\end{tabular}
\end{ruledtabular}
\end{table}

\section{Results}
\label{sec:results}
In this section results are presented for various observables 
that have been computed using an equation of state for matter 
composed of neutrons, protons, electrons, and muons in beta 
equilibrium:
\begin{subequations}
 \begin{eqnarray}
  && n \leftrightarrow p + e^{-} + \bar\nu_{e}\;, 
       \label{pnequil}\\
  && e^{-} \leftrightarrow \mu^{-} + \nu_{e} + \bar\nu_{\mu}\;.     
       \label{emequil}
 \end{eqnarray}
\end{subequations}
These reactions demand that the chemical potential ($\mu$) of the 
various constituents be related by the following equation:
\begin{figure*}[ht]
\begin{center}
\includegraphics[width=6.00in,angle=0,clip=true]{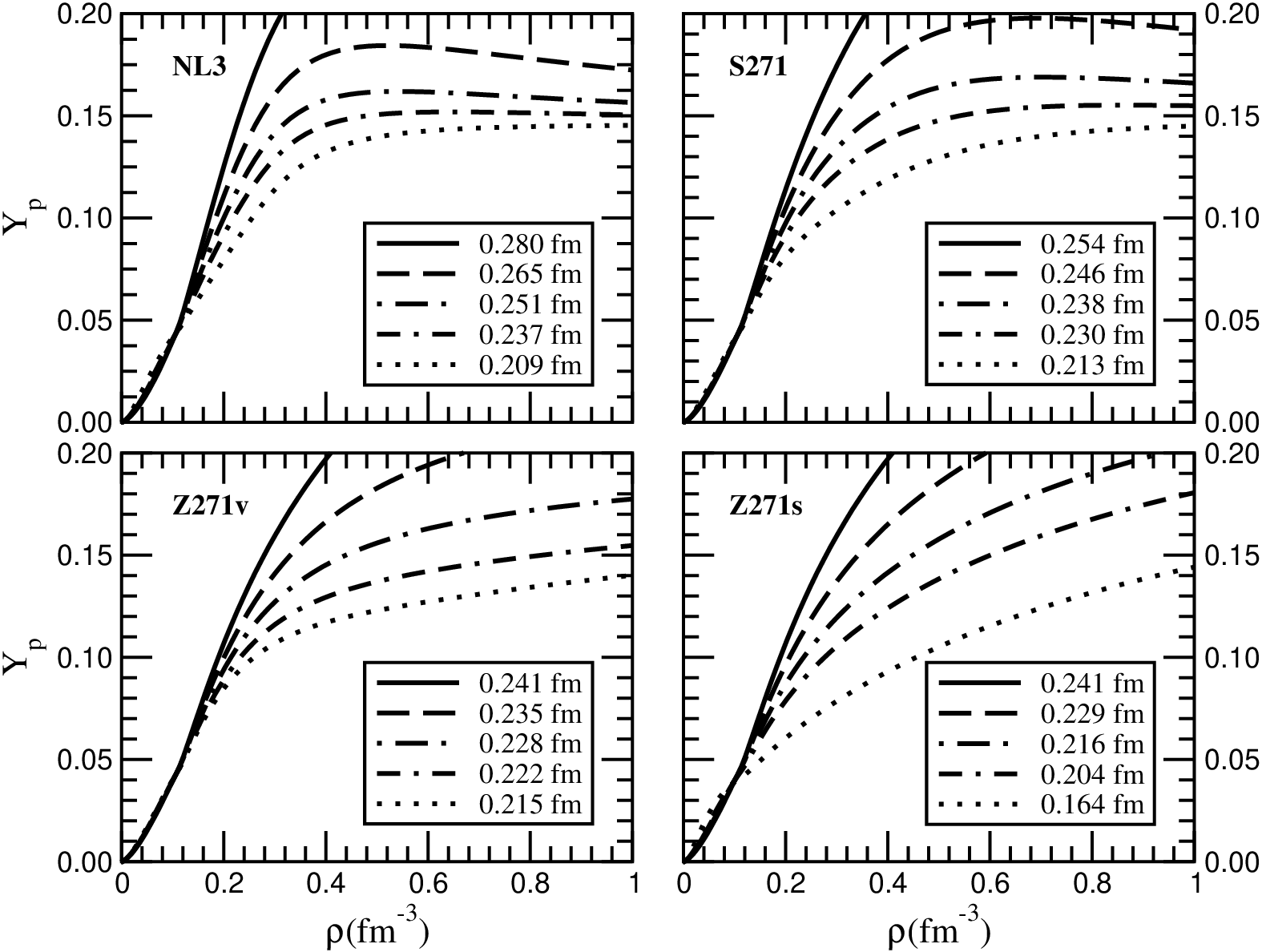}
\caption{Proton fraction $Y_p$ versus baryon density 
	 $\rho$ for neutron-rich matter in beta equilibrium 
	 for all parameter sets discussed in the text. For 
	 every given set, the curves represent different 
	 values of the isoscalar-isovector coupling
         ($\Lambda_{\rm v}$ or $\Lambda_{\rm s}$) that 
         yield the values of $R_n\!-\!R_p$ indicated in 
	 the inset. See also the various tables in the
	 text.} 
\label{Figure1}
\end{center}
\end{figure*}
\begin{equation}
 \mu_{n}-\mu_{p}=\mu_{e}=\mu_{\mu}\;,
 \label{mu1}
\end{equation}
where $\mu_n$, $\mu_p$, $\mu_e$, and $\mu_\mu$ represent the chemical
potentials of neutrons, protons, electrons, and muons, respectively.
Neglecting the rest mass of the electron, Eq.~(\ref{emequil}) is
equivalent (for $k_{\rm F}^{e}\!\ge\! m_{\mu}$) to the following 
equation expressed in terms of the Fermi momenta of the electron and 
the muon:
\begin{equation}
 k_{\rm F}^{e}=
 \sqrt{\left(k_{\rm F}^{\mu}\right)^{2}+m_{\mu}^{2}}\;.
 \label{mu2}
\end{equation}
Finally, charge neutrality imposes the following constraint on the
system: $\rho_{p}\!=\!\rho_{e}\!+\!\rho_{\mu}$, or equivalently, 
\begin{equation}
  \left(k_{\rm F}^{p}\right)^{3}=
  \left(k_{\rm F}^{e}\right)^{3}+
  \left(k_{\rm F}^{\mu}\right)^{3} \;.
 \label{charge}
\end{equation}
For a given proton Fermi momentum, the corresponding Fermi momenta
for the electrons and the muons are readily obtained by solving 
Eqs.~(\ref{mu2}) and~(\ref{charge}). With these in hand, 
Eq.~(\ref{mu1}) determines the equilibrium neutron ($Y_{n}\!=\!N/A$) 
and proton ($Y_{p}\!=\!Z/A$) fractions in the system.

In Fig.~\ref{Figure1} the proton fraction $Y_p$ for matter in beta
equilibrium is shown as a function of the baryon density for all
models discussed in the text (see Table~\ref{Table1}). The various
curves displayed in each panel are for values of $\Lambda_{\rm s}$ or
$\Lambda_{\rm v}$ that give the indicated values for the neutron skin
of $^{208}$Pb. Note that the neutron skin of a nucleus is defined as
the difference between the neutron ($R_n$) and the proton ($R_p$)
root-mean-square radii. All of the curves yield the same proton
fraction at low density ($k_{\rm F}\!=\!1.15$~fm$^{-1}$ or
$\rho\approx 0.1$~fm$^{-3}$) because the symmetry energy has been
adjusted to $a_{\rm sym}\!=\!25.67$~MeV in order to reproduce the
binding energy $^{208}$Pb. The proton fraction increases more rapidly
with density for those parameter sets that yield larger neutron radii
in $^{208}$Pb (namely, those with a stiffer symmetry energy).  Thus,
the neutron radius of $^{208}$Pb constrains the slope $dY_p/d\rho$ at
normal densities. This enables one to make a more reliable
extrapolation of $Y_p$ to the higher densities where it displays some
model dependence.

\begin{figure}[ht]
\begin{center}
\includegraphics[width=3.25in,angle=0,clip=true]{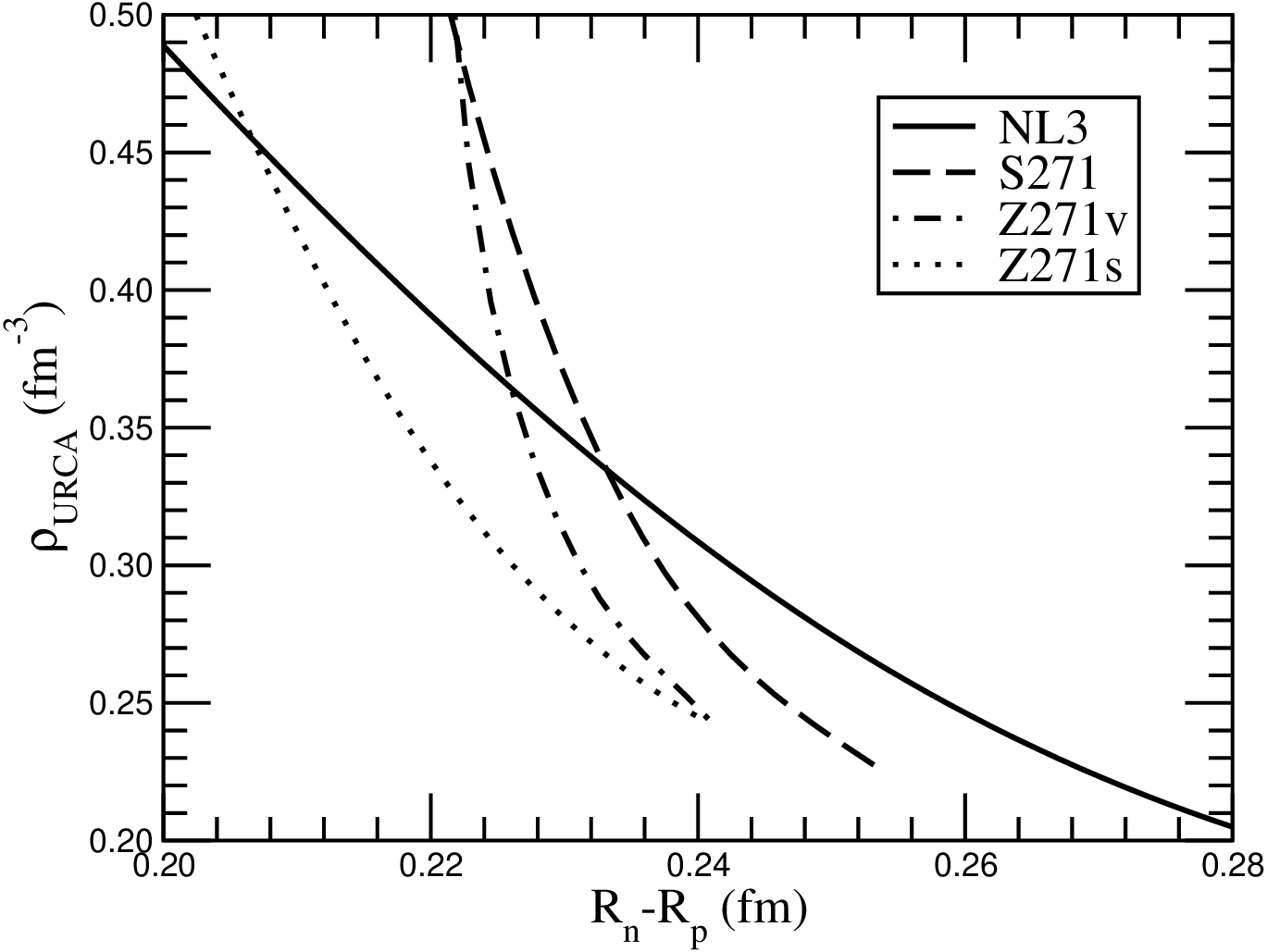}
\caption{Threshold density for the direct URCA process versus 
         the predicted neutron skin ($R_n$-$R_p$) of
	 $^{208}$Pb. Parameter sets NL3, S271, and Z271v
	 use a nonzero value for $\Lambda_{\rm v}$ while
	 Z271s uses a nonzero $\Lambda_{\rm s}$.}
\label{Figure2}
\end{center}
\end{figure}

The direct URCA process is viable only when the proton fraction is
large enough to conserve momentum in the neutron beta decay reaction
of Eq.~(\ref{urca1}). Hence, for this reaction to proceed, the Fermi
momenta of neutrons, protons, and electrons must satisfy the following
relation:
\begin{equation}
  k_{\rm F}^{n}\le k_{\rm F}^{p}+k_{\rm F}^{e}\;.
 \label{URCAkF}
\end{equation}
The URCA threshold density $\rho_{\scriptscriptstyle{\rm URCA}}$ 
is defined as the density at which the equality 
($k_{\rm F}^{n}\!=\!k_{\rm F}^{p}\!+\!k_{\rm F}^{e}$) is satisfied.  
Note that in the simplified case of matter without muons, that is, 
$k_{\rm F}^{e}\!<\!m_{\mu}$ and $k_{\rm F}^{p}\!=\!k_{\rm F}^{e}$, 
the proton fraction at the onset of the direct URCA process is
$Y_{p}\!=\!1/9\!\sim\!0.111$. In the opposite limit, 
$k_{\rm F}^{e}\!\gg\!m_{\mu}$, the threshold proton fraction is
$Y_{p}\!\sim\!0.148$. Thus, the threshold proton fraction must
be contained within these two values for all baryon densities
(see Tables~\ref{Table2}-\ref{Table5}).
In Fig.~\ref{Figure2} the URCA threshold density 
is displayed as a function of the neutron skin $R_n\!-\!R_p$ 
of $^{208}$Pb. There is a clear tendency for 
$\rho_{\scriptscriptstyle{\rm URCA}}$ to decrease with increasing
neutron skin. Recall that a large neutron skin implies a stiff 
symmetry energy and a large proton fraction. Thus the onset for 
the direct URCA process for models with large neutron skins 
occurs at low baryon densities.
Indeed, parameter sets with neutron skins of $R_n\!-\!R_p\agt0.24$~fm 
yield a relative low URCA density of
$\rho_{\scriptscriptstyle{\rm URCA}}\alt0.30$~fm$^{-3}$.
This density is only a factor of two larger than normal 
nuclear matter saturation density 
($\rho_0\approx 0.15$~fm$^{-3}$). In contrast, if 
$R_n\!-\!R_p\alt0.21$~fm, the onset for the URCA process
is above $3\rho_0$.

\begin{figure}[ht]
\begin{center}
\includegraphics[width=3.25in,angle=0,clip=true]{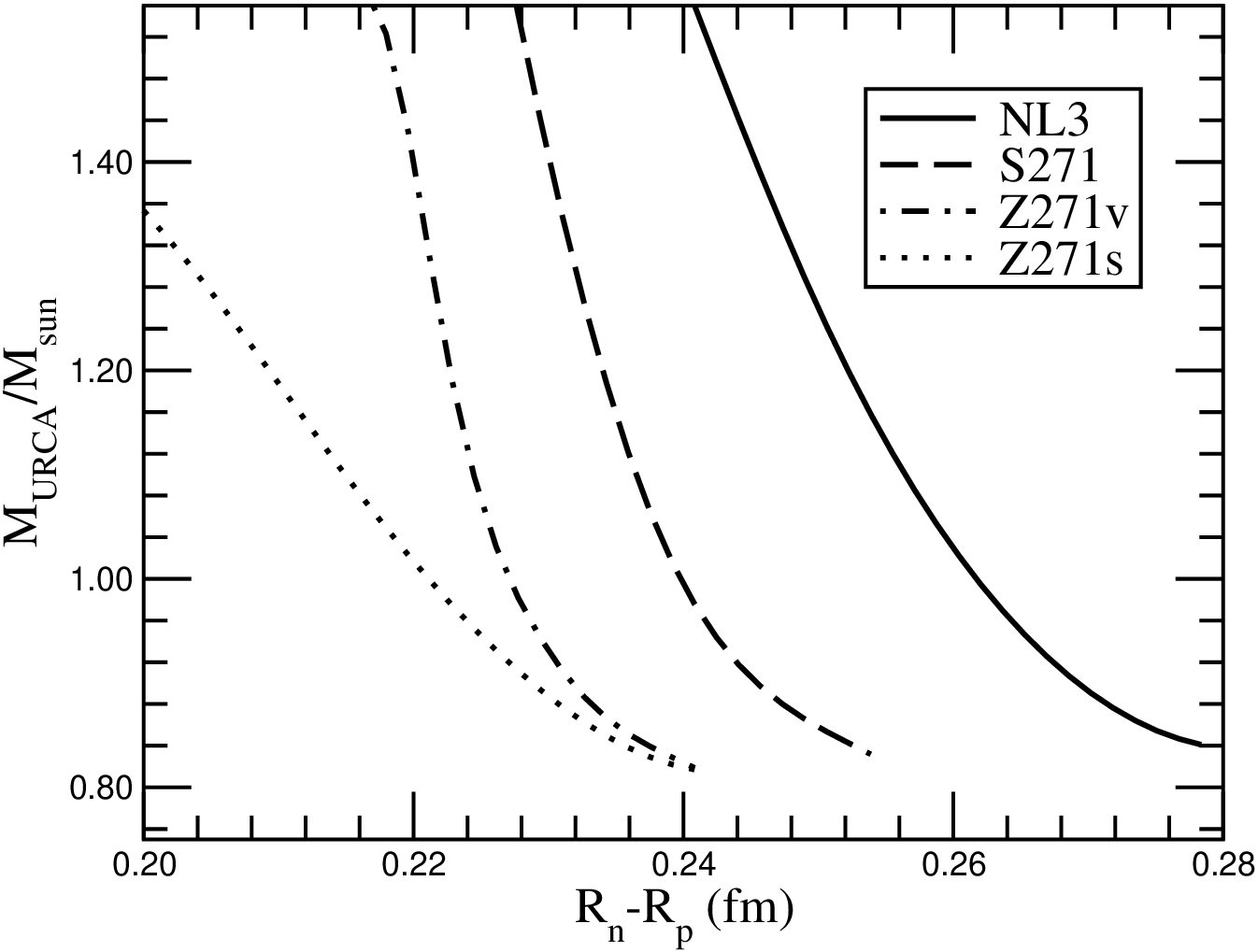}
\caption{Threshold neutron-star mass for the direct URCA process 
	 versus the predicted neutron skin ($R_n$-$R_p$) of
	 $^{208}$Pb. Parameter sets NL3, S271, and Z271v
	 use a nonzero value for $\Lambda_{\rm v}$ while
	 Z271s uses a nonzero $\Lambda_{\rm s}$.}
\label{Figure3}
\end{center}
\end{figure}

The structure of spherical neutron stars in hydrostatic equilibrium is
solely determined by the equation of state of neutron-rich matter in
beta equilibrium. Having specified the equation of state, we determine
the mass of neutron stars that may cool via the direct URCA process by
integrating the Tolman-Oppenheimer-Volkoff equations. Our treatment of
the low-density crust, where the matter in the star is nonuniform,
will be discussed in a later work~\cite{Ho02}.  This region, however,
has almost no effect on neutron-star masses. As mentioned earlier, we
consider matter composed of neutrons, protons, muons, and (massless)
electrons.  In Fig.~\ref{Figure3} we display, as a function of the
neutron skin in $^{208}$Pb, the mass of a neutron star whose central
density equals the URCA density ($\rho_{\scriptscriptstyle{\rm
URCA}}$) of Fig.~\ref{Figure2}. Neutron stars with larger masses, and
thus higher central densities, will cool by the direct URCA process;
those with lower masses will not. There is an obvious trend for this
threshold mass to decrease with increasing $R_n\!-\!R_p$. Recall that
the onset for URCA cooling in models with large neutron skins occurs
at low baryon densities, thus lower ``URCA masses''. If the neutron
skin of $^{208}$Pb is less than about $R_n\!-\!R_p\alt 0.20$~fm, then
all parameter sets considered in this work predict that a neutron star
of 1.4~$M_\odot$ will not undergo URCA cooling. Conversely, if
$R_n\!-\!R_p\agt 0.25$~fm, then all parameter sets allow URCA cooling
for 1.4~$M_\odot$ neutron star. Note that all well measured neutron
stars have masses near 1.4~$M_\odot$~\cite{Th99}.
The threshold neutron star mass for the direct URCA process, 
$M_{\rm URCA}$, depends on both the ``critical'' URCA density
$\rho_{\scriptscriptstyle{\rm URCA}}$ (of Fig.~\ref{Figure2}) 
and on the equation of state at low and high densities. Yet 
the neutron skin of $^{208}$Pb constrains only the low density 
EOS~\cite{brown}, thereby generating the model dependence 
displayed by Fig.~\ref{Figure3}. In particular, the radius 
of a neutron star, although correlated to $R_n\!-\!R_p$, is 
not determined uniquely by it~\cite{radii}. This suggests that 
models with a stiff EOS at high density, such as the NL3 parameter 
set, will yield neutron stars with relatively large radii and low 
central densities. This implies, for example, that for a fixed 
central density of
$\rho_{\scriptscriptstyle{\rm URCA}}\!=\!0.3$~fm${}^{-3}$, the 
NL3 set (with large radii) generates an URCA mass of 
$M_{\rm URCA}\!\simeq\!1.4~M_{\odot}$; in contrast, the softer 
Z271s set (with small radii) yields an URCA mass of only 
$M_{\rm URCA}\!\simeq\!1~M_{\odot}$. Equivalently, to make 
a $M_{\rm URCA}\!=\!1.4~M_{\odot}$ neutron star the NL3
set requires an interior density of
$\rho_{\scriptscriptstyle{\rm URCA}}\!=\!0.3$~fm${}^{-3}$,
while a central density almost twice as large is needed for the 
Z271s set to generate the same mass neutron star. 
These facts suggest that an accurate measurement of neutron-star 
radii may reduce the model-dependence ({\it i.e.,} the spread) 
observed in Fig.~\ref{Figure3}. Yet, even without further constraints 
the spread is relatively small and a measurement of $R_n\!-\!R_p$ in 
$^{208}$Pb may still prove decisive.

\section{Summary and Conclusions}
\label{sec:conclusions}
Recent X-ray observations suggest that some neutron stars cool
quickly. This enhanced cooling could arise from the direct URCA
process---that requires a high proton fraction---or from the beta
decay of additional hadrons in dense matter, such as pions, kaons,
hyperons, or quarks. Yet, it seems unlikely that the X-ray
observations alone will determine the origin of the enhanced cooling.

In this work we propose to use a laboratory experiment to constrain
the direct URCA process in neutron stars. The Parity Radius Experiment
at the Jefferson Laboratory~\cite{prex,bigpaper} aims to measure the
neutron radius of $^{208}$Pb accurately and model independently via
parity-violating electron scattering.  For the direct URCA process to
be realized, the equilibrium proton fraction in the star must be
large. The equilibrium proton fraction is determined by the symmetry
energy, whose density dependence can be strongly constrained through a
measurement of the neutron radius in $^{208}$Pb. Such a measurement
could provide a reliable extrapolation of the proton fraction to
higher densities. Thus, predictions for the neutron radius in
$^{208}$Pb have been correlated to the proton fraction in dense
neutron rich matter by using a wide range of relativistic
effective-field theory models.  We find that models with a neutron
skin in $^{208}$Pb of $R_n\!-\!R_p\alt0.20$~fm generate proton
fractions that are too small to allow the direct URCA process in
1.4~$M_\odot$ neutron stars. Conversely, if $R_n\!-\!R_p\agt0.25$~fm,
then all models predict the URCA cooling of 1.4~$M_\odot$ stars.

While this paper has focused on relativistic effective field-theory
models, we expect our conclusions to be general and applicable to
other approaches, both relativistic and nonrelativistic. For example,
the nonrelativistic equation of state of Friedman and
Pandharipande~\cite{fp} predicts too small a proton fraction for URCA
cooling to be possible. Moreover, this equation of state yields a
neutron skin in $^{208}$Pb of only
$R_n\!-\!R_p\!=\!0.16\pm0.02$~fm~\cite{brown}. Thus, these results
are fully consistent with Fig.~\ref{Figure3} that predicts no URCA
cooling for such a small value of $R_n\!-\!R_p$.

The equation of state considered in this work consists of matter
composed of neutrons, protons, electrons, and muons in beta
equilibrium; no exotic component was invoked. Further, no explicit
proton or neutron pairing was considered. Nucleon superfluidity is 
an accepted phenomenon in nuclear physics and superfluid gaps are 
important for the cooling of neutron stars~\cite{yakovlev}. Thus,
the study of pairing gaps in relativistic effective-field theories 
is an important area of future work; first steps in this direction 
have been taken in Ref.~\cite{super}. In particular, the proton 
pairing gap in matter with a high proton concentration must be 
computed~\cite{t5}.

In summary, the feasibility of enhanced cooling of neutron stars via
the direct URCA process was studied by correlating the proton fraction
in dense, neutron-rich matter to the neutron skin of $^{208}$Pb. Thus,
a measurement of the neutron radius in $^{208}$Pb may become vital for
confirming (or dismissing) the direct URCA cooling of neutron
stars. If direct URCA cooling is ruled out, then observations of
enhanced cooling may provide strong evidence in support of exotic
states of matter, such as meson condensates and quark matter, at the
core of neutron stars.

\begin{acknowledgments}
This work was supported by the U.S.~DOE
under contracts DE-FG02-87ER40365 and DE-FG05-92ER40750.
\end{acknowledgments}

\end{document}